# Managing Level of Detail through Head-Tracked Peripheral Degradation: A Model and Resulting Design Principles

*Benjamin Watson, Neff Walker & Larry F. Hodges*

Graphics, Visualization and Usability Center

Georgia Institute of Technology

801 Atlantic Dr, Atlanta, GA 30332-0280 USA

tel: +1 404 894 9389; Email: watsonb@cc.gatech.edu

## ABSTRACT

Previous work has demonstrated the utility of reductions in the level of detail (LOD) in the periphery of head-tracked, large field of view displays. This paper provides a psychophysically based model, centered around an eye/head movement tradeoff, that explains the effectiveness of peripheral degradation and suggests how peripherally degraded displays should be designed. An experiment evaluating the effect on search performance of the shape and area of the high detail central area (inset) in peripherally degraded displays was performed, results indicated that inset shape is not a significant factor in performance. Inset area, however, was significant: performance with displays subtending at least 30 degrees of horizontal and vertical angle was not significantly different from performance with an undegraded display. These results agreed with the proposed model.

## 1 INTRODUCTION

As virtual environments (VE) researchers attempt to broaden the range of applications for VE technology, they are attempting to display ever larger and more complex models. Many of these models, however, cannot be displayed with acceptable frame rates in current systems. Several researchers have identified this "frame" or "update rate" problem as one of the most pressing facing the VE community [NSF, 1992; Van Dam, 1993]. Foremost among the proposed solutions to this problem is the idea of varying *level of detail* (LOD). As used by most VE researchers, this phrase refers to model and rendering complexity, which can be varied to ensure that VEs are rendered at some acceptable frame rate.

Although the search for solutions to the frame rate problem has spawned a large and highly technical body of research, the problem at its heart is essentially a human one. The core concern is inadequate frame rate and system responsiveness, and the resulting impact on human performance. LOD solutions reduce displayed detail in order to improve system responsiveness, instituting a tradeoff between visual and temporal fidelity. Visual fidelity, however, can also impact human performance. Thus designers of VE or other interactive 3D graphics displays must balance the human performance impact of changed system responsiveness against the impact of the corresponding change in displayed visual detail. We call this tradeoff the *LOD tradeoff* and refer to the management of this tradeoff with the term *LOD management*.

There are many techniques that might be used to vary displayed detail, including using geometric models of varying degrees of accuracy [DeRose & Lounsberry, 1993; Rossignac & Borrel, 1992; Turk, 1992; Varshney, Agarwal, Brooks, Wright, & Weber, 1995], lighting and shading models of differing levels of realism, and textures and graphics windows of differing resolution [Maciel & Shirley, 1995]. Many researchers have compared the relative impact on human performance of different graphics rendering techniques in traditional display environments [Atherton & Caporeal, 1985; Barfield, Sandford, & Foley, 1988; Booth, Bryden, Cowan, Morgan, & Plante, 1987]. In general, these studies showed significant effects on performance when rendering method is varied. However, in most cases a point of diminishing returns was reached, beyond which additional image complexity and computation produced insignificant improvement in user performance. This suggests that varying LOD by using different rendering techniques may be a promising LOD management approach. However, we are not aware of any studies beyond our own that address the impact on human performance of varying detail by using models of varying accuracy.

Many LOD management techniques attempt to sidestep the LOD tradeoff by reducing detail only when it is not perceivable, giving the benefits of improvements in system responsiveness without the corresponding cost of detail loss. There are two existing examples of this approach. The first takes advantage of the eye's reduced ability to perceive objects as the size of those objects decreases. Flight simulators and VE systems [Funkhauser & Sequin; 1993, Maciel & Shirley, 1995] exploit this characteristic by using lower LODs when the visual angle of a portion of the model decreases. This technique has a proven track record in the flight simulator industry. The second example takes advantage of the eye's reduced ability to perceive objects out of the center of the field of view [Bishop, 1986]. This characteristic of the eye suggests a divided display containing a central, high detail inset, corresponding to the center of the eye's field of view; as well as a surrounding, simpler periphery, corresponding to the peripheral areas of the field of view [e.g., Reddy, Watson, Walker, & Hodges, in press]. Funkhauser, Sequin, and Maciel & Shirley [1995] have implemented systems that degrade peripheral detail to improve frame rate.

This *peripheral degradation* technique requires a knowledge of the current gaze direction, so that the high detail inset may placed in the center of the field of view. Ideally, in order to track gaze position, both head and eye movement should be tracked. There are a number of reasons, however, to avoid eye tracking when possible. Eye tracking devices are currently expensive. They can be a bulky and intrusive part of an interface, especially in large screen environments that allow significant amounts of user movement. Eye trackers generally require recalibration for each user. Finally, as an additional input into the rendering loop, eye tracking can reduce system responsiveness.

In [Watson, Walker, Hodges & Worden, 1995; 1996], we presented a study that demonstrated the effectiveness of peripheral degradation when used with head tracking alone. The study showed that detail could often be cut by one half without significantly affecting user performance in a highly demanding search task. This demonstrated the ìsidestepî of the LOD tradeoff mentioned above, allowing application designers to use this technique to raise frame rate without concern for the impact of the corresponding loss in detail. Results showed that severe reductions in peripheral detail were possible with insets subtending at least 22 degrees of horizontal angle.

In this paper, we propose a model outlining the reasons for the effectiveness of head-tracked peripheral degradation. This model is based on the psychophysical parameters of head and eye movements. From this model, we derive certain principles for the use of the peripheral degradation technique. We then describe an experiment that verifies certain model predictions, and that addresses the issue of the shape of the high detail inset.

## 2.1 Retinal Physiology and Eye-Head Coordination

Psychophysical research has shown that visual search is a series of fixations. The center of the retina, called the fovea, is used for a high detail search in the area of fixation, while the periphery of the retina is used during selection of the next fixation point. Visual acuity drops exponentially outside the fovea, making visual display of a level of detail adequate to guide selection of the next fixation quite easy.

Study of simultaneous horizontal eye and head motion with the rest of the body fixed [Barnes, 1979] has shown that:

- The eye can accelerate to reach a target fixation point much more quickly than the head.

- Fixation will often occur without head motion when the target is offset from the current line of sight by 30 degrees of visual angle or less, with the likelyhood of head motion increasing as offset angle increases.

- Head and eye motion begin almost simultaneously when the target fixation is offset by at least 15 degrees of visual angle.

- The eye reaches the target fixation point before the head does, as the head reaches the target, the eye reverses direction and moves back towards a centered position relative to the head.

- Even after gaze is stable on a new fixation, the eye is still not fully centered relative to the head: it can be offset by as much as 15 degrees.

## 2.2 A Qualitative Model For Human Performance With Peripheral Degradation

Given a peripheral level of detail adequate to guide search, the selection of fixation points will not change, only the speed with which they are reached. Thus the use of the peripheral degradation technique without eye tracking may harm performance only by requiring head motion to reach the fixation point when eye motion would have been used in an undegraded display. With the knowledge about eye-head coordination presented above, we would predict the following:

- Insets subtending less than 30 degrees of horizontal angle can harm visual search performance.

- Insets subtending more than 30 degrees of horizontal angle will not improve performance.

The sparseness of research on vertical eye and head motion did not allow development of an apriori model of the effects of vertical inset extent on search performance, nor of the effects of inset shape.

## 3 EXPERIMENT

We designed an experiment that would provide data to expand the predictive power of our peripheral degradation model to account for vertical and shape inset parameters. The experiment was also designed to allow verification of the a priori horizontal elements of our model.

## 3.1 Experimental Methodology

3.1.1 *Participants*. Eight graduate students participated in the experiment. All reported normal vision, none wore eyeglasses.

3.1.2 *Design*. The study utilized a two factor design. The primary independent variables were within-subjects variables. These variables were horizontal and vertical inset extent (10, 20, 30 or 40 degrees). The study also had two control variables. These variables were trial condition (target object was either present or absent) and number of trials (6 target absent conditions, 14 target present). Both of these control variables were within-subjects. In addition to these conditions, we also had subjects run in an undegraded (normal) display as a check on baseline performance. The performance in this condition will be analyzed separately.

3.1.3 *Apparatus*. Subjects wore a Virtual Research Flight Helmet [Robinett & Rolland, 1992] to immerse themselves in the experimental environment. The Flight Helmet mounts two color LCD displays on the user's head, each with vertical field of view of 58.4 degrees, and a horizontal FOV of 75.3 degrees. Each LCD contains an array of 208 x 139 color triads. We used the Flight Helmet in a monoscopic, biocular mode by sending the same image to each of the video inputs, and mounting plastic fresnel lenses on the HMD optics to remove interocular disparity. Average horizontal angular resolution is 21.74 arcmins.

The motion of a subject's head in the Flight Helmet was tracked with the Polhemus Isotrak II 3D tracking hardware. The monoscopic images sent to the Flight Helmet were generated by a Silicon Graphics Crimson Reality Engine, using the gl graphics library and the SVE virtual environments library [Kessler, 1993]. Silicon Graphics' scan converting hardware and software were used to convert these images into an NTSC signal. Subjects used a plastic mouse shaped like a pistol grip to respond to the experimental environment. The mouse had three buttons for the thumb mounted on top, and one button for the index finger mounted on the front. The mouse was not tracked. When using the experimental environment, subjects stood inside a 4í diameter platform raised six inches and surrounded by a three foot railing.

The virtual experimental environment consisted of a floor,

indicated by a grid of white lines on black. The background above the floor was also black. A home object indicated starting position for search, and was textured with a bullseye design. Users were not represented in the virtual environment.

3.1.4 *Procedures*. During each trial, subjects would search through five randomly located, identically sized objects for a target object. Most objects were textured with identical images of a smiling human face. The single target object was textured with the same face, with the mouth closed. Objects always appeared at the same virtual distance, and subtended a horizontal visual angle of approximately 12 degrees. The mouth subtended only three degrees (Figure 1).

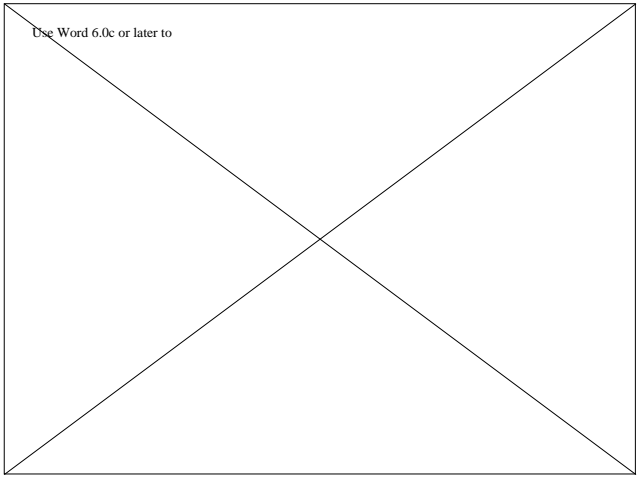

**Figure 1:** Display view with 40° x 30° high detail inset. Subjects located the face with its mouth closed.

LOD was reduced in the periphery by changing resolution. Two views of the experimental environment were generated, a high detail one containing 208 x 139 pixels (the same as the HMD), and a low detail one containing 42 x 28 pixels. These two views were textured onto two polygons: low detail for the periphery, and high detail for the inset. The two polygons were overlapped slightly and blended with alpha transparency to make the boundary of the high detail inset harder to detect. Texturing was done in real time with the fbsubtexload command and FAST_DEFINE [SGI, 1993]. Since eye tracking was not used, insets were always located in the center of the displayed image. Insets were rectangular, and varied in shape and size with the independent variables.

Each experimental trial consisted of a single search task. After focusing on the home object, subjects pressed a button to begin the task. After a random (between .1 and .8 seconds) delay, the home object disappeared, and objects appeared to the right of the subject's initial view. Subjects attempted to locate the target object and pressed either the thumb or index button to indicate if it was present. All objects then disappeared and the home object reappeared, with on screen feedback indicating correctness and time of response. When the subjects had again focused on the home object and pressed the appropriate button, a new search task began. In trials with a target object, subjects were not credited with a correct trial unless they had brought the target object into their view. In trials without a target object, subjects were not credited with a correct trial unless they had brought every object into their view. This forced an exhaustive search. Subjects performed the search task with each of the 17 display types mentioned above.

Several other variables were controlled. Maximum frame rate was 30, and frame rates did not fall below 29.9. Objects were distributed in a search space equal to four HMD fields of view, totalling 150 horizontal degrees and 118 vertical degrees. Our previous experiment [Watson et al., 1996] showed that sparseness of object distribution, location of the objects relative to initial head position, and number of objects did not interact with inset size. Therefore these variables were held constant: objects were randomly clustered within a single HMD field of view, the cluster was located randomly within the search space, and there were always five objects within the cluster.

Subjects performed a total of 340 correct trials over all displays, and 20 correct trials with each display before moving on to the next. The order in which different displays were presented was randomly varied between subjects, and controlled so that no two display sequence was used more than three times across subjects.

Each subject completed the experiment in a single day. At the beginning of the experiment, subjects were told the nature of the experiment, and given 20 practice trials to ensure that they understood the general nature of the task. At the beginning of each 20 trial display block, subjects performed five practice trials with the display. None of these practice trials were included in the analyses. For any trial on which the subject made an error, that trial was repeated by placing it back into the queue of remaining trials.

## 3.2 Results

The data from the visual search task were analyzed by means of four factor analyses of variance. We report all significant effects that have a probability level of 0.05 or less. Bonferroni pair-wise comparisons were used to follow up significant main effects and simple main effects. The independent variables were horizontal and vertical inset size. The dependent variables were mean search time when the target was present, and accuracy when target was present. For the two dependent variables, the means are based on collapsing across trials. The means for these two measures by horizontal and vertical extent of the high detail display are presented in Table 1 and Figure 2. Initially we performed two way ANOVAs (horizontal and vertical extent) excluding the undegraded condition. Performance in that display condition was compared to the inset conditions in a separate analysis.

The 2 X 2 analysis of variance on search time when the target was present revealed two significant main effects. There was a significant main effects for horizontal extent of the inset [F(3,21) = 8.84, p < .001]. Follow-up analyses showed that the search times were longer when the inset's horizontal extent was 10 degrees than when it was 30 or 40 degrees. Search times with insets of 20 degrees did not differ significantly from those when the inset size was 10, 30 and 40 degrees. No other pair-wise comparisons were significant.

There were also significant main effects of vertical extent of the high detail inset [F(3,21) = 6.16, p < .001]. Search times were faster when the vertical extent of the inset was 30 or 40 degrees than when the inset extent was 10 degrees. Again, no other pair-wise comparisons were significant.

One final analysis was performed comparing mean search time to the 40 degrees horizontal and 40 degrees vertical extent inset to the full high detail display (75 horizontal and 58 degrees vertical extent). A one way analysis of variance was performed comparing search time in these two displays. This analysis failed to reveal any differences in search time between these two display conditions.

**Table 1:** Mean search times in seconds and percentage trials correct for differently sized high detail areas in the target present conditions. Rows have the same horizontal extent, columns the same vertical extent. Mean search time for the undegraded 75° x 58° display was 2.85 seconds, mean accuracy was 94.9%.

| H \ V | 10° | 20° | 30° | 40° |
|---|---|---|---|---|
| 10° | 4.147 92.1% | 4.150 97.6% | 4.058 96.5% | 3.538 95.4% |
| 20° | 3.721 97.3% | 3.552 93.3% | 3.398 98.2% | 3.448 99.1% |
| 30° | 3.601 97.3% | 3.451 95.9% | 2.876 94.6% | 3.147 94.1% |
| 40° | 3.808 99.1% | 3.105 95.8% | 3.281 92.6% | 3.061 95.8% |

The 4 X 4 analysis of variance on accuracy when the target was present revealed only one significant effect, a significant interaction of horizontal and vertical extent [$F_{(9,63)} = 2.67$, $p < .05$]. Neither main effect approached significance. Follow-up analyses revealed that accuracy seemed to be most strongly affected when both visual extents were 10 degrees. It appears that area of the inset was more important than the either of the two dimensions. We would also like to point out that in all display conditions, accuracy was over 92% and several subject reported that their errors were due primarily to ìslipsî, where they pressed the wrong button, rather than true errors of failing to see the target. Therefore inset size seemed to have little effect on accuracy in our task.

## 4 CONCLUSION

The results of this experiment confirm and extend the results of our earlier work [Watson, et al, 1996]. In this experiment we have validated our argument that detail in the periphery of a head-tracked display can be reduced without affecting search performance. Based on our previous work and on work on head and gaze coordination [Barnes, 1979], we had hypothesized that the only cost to having only a central area of high visual detail would be that fast and accurate eye movements would be replaced by slower head movements. This tradeoff can occur anytime that detail was lowered within 45 horizontal degrees of the fixation point, as 45 degrees is the physical extent of eye movement. However, previous work had suggested that while eye movements alone can extend 45 degrees, outside of a ±15 degree area around the initial fixation point, a combination of head and eye movements are normally used when moving to a new fixation point. Based on this, in comparison to an undegraded normal display, we anticipated slower search times when the high detail inset subtended less than 30 horizontal degrees, and similar search times when the inset subtended 30 horizontal degrees or more. This is the pattern of results that we found in this experiment. In addition, we determined that the 30 degrees cutoff value applied to the vertical extent of the inset. We found no effects of inset shape.

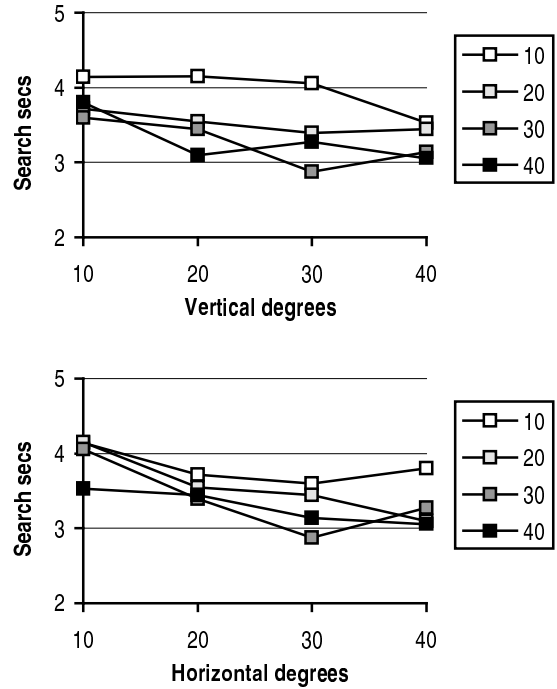

**Figure 2:** Two plots of the data from Table 1. In the upper plot, the abscissa indicates vertical inset extent, color horizontal extent. In the lower plot, the abscissa indicates horizontal extent, color vertical extent.

These results suggest that LOD management with head-tracked peripheral degradation is an extremely useful technique. With the display used in this experiment, we were able to reduce detail over 80% of the display area to extremely low levels (containing only 42 x 28 pixels) without harming visual search performance. As display field of view increases, the benefits of this approach become even more salient. For example, in a CAVE environment with a 270 by 270 degree display area, a 30 by 30 degree inset would take up roughly 1.25% of available display space. The total amount of savings, of course, would depend on the manner in which detail is reduced in the periphery.

The computational savings that can be gained from this technique have three obvious applications that can be used to improve performance in a virtual environment. One approach is to utilize these saving to increase system responsiveness. In this manner, frame rate can be increased, lag decreased and resulting user performance improved. A second approach would be to maintain current system responsiveness while increasing the field of view provided by the display. In many tasks, particularly search, wider fields of view aid performance. Finally, one could use the computational savings to present an even higher level of visual detail within the inset.